# Line geometry and electromagnetism III: groups of transformations

D. H. Delphenich
Kettering, OH 45440

**Abstract.** The role of linear and projective groups of transformations in line geometry and electromagnetism is examined in accordance with Klein's Erlanger Programm for geometries. The group of collineations of real projective space is chosen as the most general group and reductions to some of its various subgroups are then detailed according to their relevance to electromagnetic fields, and especially wave-like ones.

**Table of contents**

**1. Introduction.** This paper is the third in a series of papers (cf., [**1**, **2**]) that examine the role of line geometry in the theory of electromagnetism. Hence, we shall try to minimize the overlap in the presentation, although it shall occasionally be advantageous to recall some of the relevant facts that were stated previously.

As we have shown, the main point of contact between the two theories is the Plücker-Klein embedding that allows one to represent lines in $\mathbb{R}P^3$ as decomposable bivectors **a** ^ **b** (2-forms $\alpha \wedge \beta$, resp.) on $\mathbb{R}^4$ – or rather, the lines [**a** ^ **b**] ([$\alpha \wedge \beta$], resp.) through the origins of the relevant vector spaces, which we denote by $\Lambda_2$ ($\Lambda^2$, resp.). The condition that a bivector **B** (2-form $B$, resp.) should be decomposable is an algebraic and homogeneous quadratic one, namely that **B** ^ **B** = 0 ($B \wedge B = 0$, resp.), and the quadric hypersurface $\mathcal{K}$ ($\mathcal{K}^*$, resp.) that it defines in $P\Lambda_2$ ($P\Lambda^2$, resp.) is called the "Klein quadric."

One then gets one-to-one correspondences between the points of $\mathcal{L}(3)$ (viz., the set of all lines in $\mathbb{R}P^3$) and $\mathcal{K}$ or $\mathcal{K}^*$.

The relevance of bivector fields and 2-forms to the theory of electromagnetism has been established since at least the time of Élie Cartan, who also commented on the fact that the Lorentzian structure of spacetime is only used in Maxwell's equations implicitly by way of the Hodge duality operator *. That comment of his echoed a more extensive exposition of that fact by Kottler, and it was developed further by David van Dantzig, Friedrich Hehl, Yuri Obukhov, the author, and several others ([1]).

Previously, we discussed the basic associations of electromagnetic concepts with line-geometric ones and the fact that the existence of electromagnetic waves as solutions of the pre-metric Maxwell equations implies the existence of a dispersion law for those waves. Indeed, for nonlinear electromagnetic waves, there would generally be a different dispersion laws for each "type" of wave-like solution, so we first addressed the linear case, which already gives one much to ponder in regard to line geometry.

In this installment of the series, we are going to approach the subject of how electromagnetism relates to line geometry in the spirit of Felix Klein's Erlanger Programm [**3**]. In that celebrated Habilitationsschrift, which he presented as his Inaugural lecture upon accepting a position in the mathematics department of the University of Erlangen, he characterized a "geometry" as a set of geometric objects (e.g., points, lines, hyperplanes, spheres, etc.) that were associated with some fundamental relationship (e.g., incidence, parallelism, distance, angle, etc.) and that one could then classify geometries by the groups of transformations of those geometric objects that preserved the fundamental relationship. A fundamental problem of any geometry would then be the search for invariants of those transformation groups, which could take the form of functions, numbers, hypersurfaces, or algebraic structures.

In that inaugural lecture, Klein went on to show that in his way of thinking the "Ur-geometry" that spawned the other ones was projective geometry, which Cayley phrased as "All geometry is projective geometry." The associated group of transformations would have to preserve incidence, and would constitute the group $PGL(n, \mathbb{K})$ of projective transformations for the relevant dimension. (Here, we are using $\mathbb{K}$ to generically refer to a field of characteristic zero, and usually $\mathbb{R}$ or $\mathbb{C}$.)

The groups of transformations for the lower-ranking geometries, such as affine, metric-projective, and metric-affine geometry, would then be subgroups of $PGL(n, \mathbb{K})$. For instance, one can reduce from projective to affine geometry by restricting oneself to projective transformations that preserve the (projective) hyperplane at infinity, and the relevant subgroup will be the affine group $A(n-1, \mathbb{K})$, which preserves parallelism of lines. One could also introduce a projective metric on $\mathbb{K}P^n$ and reduce to the subgroup that preserves the metric, and its algebraic structure would generally depend upon the signature type of the metric. One could also define both an affine subspace of $\mathbb{K}P^n$ and a metric on that subspace, and one would be back to the linear orthogonal spaces that one uses as the local model for Riemannian and Lorentzian manifolds (such as spacetime) by way of the tangent spaces. Thus, the Lorentzian geometry of spacetime can be seen as a reduction of a more general geometry that is mostly related to the fact the Lorentzian metric is a specialized dispersion law for the propagation of electromagnetism waves that one derives from a simplified picture of the electromagnetic constitutive properties of spacetime, namely, the "classical vacuum" assumption that amounts to making spacetime electromagnetically linear, isotropic, homogeneous, and non-dispersive (in a different sense of the word "dispersion."). However, one of the recurring themes of all quantum electrodynamics is the ubiquitous role of vacuum polarization, which suggests that the Lorentzian structure might break down long before one gets to the "Planck scale," namely, at the scale at which vacuum birefringence sets in and changes the quadratic light cone into something more quartic. This breakdown is, presumably, associated with the cloud of vacuum polarization that surrounds a point charge (or perhaps defines its true charge distribution) that charge renormalization implies.

---

([1]) For the literature of pre-metric electromagnetism, one should confer the previous articles in this series.

Hence, the ongoing theme of these articles has been that line geometry is to electromagnetism what metric geometry is to gravitation, and that in fact the appearance of gravitation becomes a sort of corollary to the electromagnetic structure of spacetime, which might, in some way, explain why gravitation is so much weaker than electromagnetism. Furthermore, although the ultimate objective is to deal with nonlinear and quantum electromagnetism, we have decided that first of all there is a lot to be established in the linear case, as well.

In the rest of this article, we shall first establish the concept of collineations, which consist of both projective transformations and correlations in the case of three dimensions. We will then show how they get represented in the invertible transformations of $\Lambda_2$ that preserve the Klein quadric. At that point, we will discuss some of the physical considerations that define reductions of the group of collineations, such as constitutive laws and dispersion laws. We will then show that the introduction of an almost-complex structure on $\Lambda_2$, which is conformally equivalent to a Lorentzian metric, first brings about a reduction of the group of collineations to the general complex linear group in dimension three and then to the special complex orthogonal group in that dimension, which is (two-to-one) isomorphic to the proper, orthochronous Lorentz group. Finally, we shall make a few remarks regarding the way that contact transformations relate to wave motion more generally, and thus, to electromagnetic waves in particular.

**2. Projective transformations.** Just as linear maps between linear spaces are defined to preserve the linear structure, one would expect a projective map to preserve a "projective structure," in some sense of the term. In particular, projective transformations must preserve the incidence of projective subspaces, where a projective subspace $A$ of $\mathbb{R}P^n$ is *incident* with a projective subspace $B$ iff either $A$ is a subspace of $B$ or $B$ is a subspace of $A$. Thus, incidence is essentially a symmetrization of the partial ordering of inclusion of projective subspaces.

This partial ordering of subspaces can be given the structure of a *lattice* by defining two binary operations on it. The first one amounts to the least upper bound of two subspaces *A* and *B*, which is called the *join* of *A* and *B* and will be written $A \vee B$; it is the smallest subspace that contains both *A* and *B*. For instance, the join of two distinct points is a line, the join of a line and point not incident on that line will be a plane, etc. The other binary operation is essentially the greatest lower bound of *A* and *B*, which is called the *meet* of *A* and *B* and will be written $A \wedge B$; it is simply the intersection of the spaces – i.e., the largest subspace that is contained in both of them.

We shall define a projective structure to be precisely this lattice of projective subspaces, when given the operations of meet and join. Hence, a *projective transformation* of $\mathbb{R}P^n$ will be an invertible map from $\mathbb{R}P^n$ to itself that preserves the lattice of projective subspaces: i.e.:

1. If *A* is a subspace of *B* then the image of *A* will be a subspace of the image of *B*; hence, the image of incident subspaces will be incident subspaces.
2. The image of a join is a join.
3. The image of a meet is a meet.

There is another kind of map that takes preserves the lattice in a "dual" sense that is called a "correlation." We shall discuss that concept in the next section.

One sees that since projective transformations take distinct points to distinct points and joins to joins, they will always take lines to lines, and planes to planes. That knowledge will suffice for $\mathbb{R}P^3$, which will be the main object of scrutiny in this study.

As it turns out, since a projective transformation does not take lines to curves, it will be covered by an invertible linear map of $\mathbb{R}^{n+1}$ to itself that is, however, defined only up to a non-zero scalar multiple; i.e.:

$$\rho \, \bar{x}^i = A^i_j \, x^j \qquad (\rho \neq 0, \det A \neq 0). \qquad (2.1)$$

Thus, one has an equivalence relation that is defined on elements of $GL(n + 1; \mathbb{R})$ that makes them equivalent when they lie along the same line through the origin in the vector space $M(n + 1; \mathbb{R})$ of all $n + 1$ by $n + 1$ real matrices. The quotient space of $GL(n + 1; \mathbb{R})$ / $\rho I$, is again a group, since $\rho I$ is the center of $GL(n + 1; \mathbb{R})$, and thus normal, and one denotes that group by $PGL(n; \mathbb{R})$ and calls it the *projective general linear group* in dimension *n*. Because $GL(n + 1; \mathbb{R})$ is $(n + 1)^2$-dimensional (as a Lie group) and $\rho I$ is one-dimensional, $PGL(n; \mathbb{R})$ will be $((n + 1)^2 - 1)$-dimensional; in particular, for $n = 3$, this dimension will be 15.

Now, any $n + 1$ by $n + 1$ real matrix *A* with a positive determinant can be expressed as a scalar multiple of a matrix $A_0 \in SL(n + 1; \mathbb{R})$ (so $\det A_0 = 1$) by way of:

$$A = (\det A)^{1/(n + 1)} A_0 . \tag{2.2}$$

For $n$ even, $n + 1$ will be odd, and the sign of $(\det A)^{1/(n + 1)}$ will be unique, but when $n$ is odd, so $n + 1$ is even, $(\det A)^{1/(n + 1)}$ might be positive or negative. Hence:

$$PGL(n; \mathbb{R}) \cong \begin{cases} SL(n+1; \mathbb{R}) & n \text{ even,} \\ SL(n+1; \mathbb{R})/\mathbb{Z}_2 & n \text{ odd.} \end{cases}$$

For the present purposes, $n$ is three, and $PGL(3; \mathbb{R}) \cong SL(4; \mathbb{R}) / \mathbb{Z}_2$.

**3. Correlations.** A *correlation* is defined to be an invertible map $[C] : \mathbb{R}P^n \to \mathbb{R}P^{n*}$. Thus, it will take points to hyperplanes, lines to subspaces of codimension 2, and so forth. It will also invert the partial ordering of inclusion so if $A$ is a subspace of $B$ in $\mathbb{R}P^n$ then $[C](B)$ will be a subspace of $[C](A)$ in $\mathbb{R}P^{n*}$. However, when one symmetrizes that relationship, one will see that correlations also preserve incidence, along with projective transformations; it will also switch meets with joins. We then summarize its effect on the projective structure by saying:
1. It inverts subspace inclusions, but still preserves incidence.
2. The image of a join will be a meet.
3. The image of a meet will be a join.

For the case of $n = 3$, one sees that a subspace of codimension two will be of dimension one. Hence, in that dimension, a correlation will also take lines to lines.

If one fixes a correlation $[C]$ then any other correlation $[C']$ will be related to it by a unique projective transformation $[L]$ of $\mathbb{R}P^n$, namely:

$$[C'] = [C][L] \qquad ([L] \equiv [C'][C^{-1}]). \tag{3.1}$$

Hence, the set of all correlations is in one-to-one correspondence with $PGL(n; \mathbb{R})$, in much the same way that $GL(n; \mathbb{R})$ divides into two diffeomorphic components according to the sign of the determinant.

A correlation $[C] : \mathbb{R}P^n \to \mathbb{R}P^{n*}$ is covered by a projective class of invertible linear maps $C : \mathbb{R}^{n+1} \to \mathbb{R}^{n+1*}$. Hence, if one fixes a basis for $\mathbb{R}^{n+1}$ and uses its reciprocal basis for $\mathbb{R}^{n+1*}$ then the correlation $[C]$ can be expressed as a system of linear equations:

$$\rho \bar{x}_i = C_{ij} x^j \qquad (\rho \neq 0, \det C \neq 0). \tag{3.2}$$

Since the image of a point [**v**] in $\mathbb{R}P^n$ under a correlation [$C$] is a hyperplane [$C(\mathbf{v})$] in $\mathbb{R}P^n$, the immediate question to ask is whether the point is incident upon the hyperplane. This comes down to the question of whether the linear functional $C(\mathbf{v})$ in $\mathbb{R}^{n+1*}$ that represents the hyperplane gives zero when applied to the vector **v**:

$$C(\mathbf{v})(\mathbf{v}) = 0. \tag{3.3}$$

One can use any correlation $C$ on $\mathbb{R}^{n+1}$ to define a non-degenerate bilinear form on $\mathbb{R}^{n+1}$:

$$C(\mathbf{v}, \mathbf{w}) = C(\mathbf{v})(\mathbf{w}). \tag{3.4}$$

Its component matrix $C_{ij}$ for any frame on $\mathbb{R}^{n+1}$ will be the same as the component matrix of the correlation itself when one uses the reciprocal frame on $\mathbb{R}^{n+1*}$.

The general correlation will have no symmetry in the permutation of **v** and **w**, but when it is symmetric, one says it defines a *polarity* on $\mathbb{R}P^n$, and when it is anti-symmetric, one says it defines an *anti-polarity*. Thus, a polarity on $\mathbb{R}P^n$ is covered by a scalar product on $\mathbb{R}^{n+1}$, and vice versa. One sees that any anti-polarity will give (3.3), but a polarity will generally satisfy that condition only for special **v**, which one calls *isotropic*.

There is a distinguished matrix − namely, $\delta_{ij}$ − that represents a correlation on $\mathbb{R}P^n$ in a canonical sort of way (at least, when one chooses a basis for $\mathbb{R}^{n+1}$). One can see that it is the matrix of the linear isomorphism of $\mathbb{R}^{n+1}$ with $\mathbb{R}^{n+1*}$ that takes the canonical frame to its reciprocal frame. Thus, any other matrix $C_{ij}$ that represents a correlation can then be expressed in the form $\delta_{ik} C^k_j$, where the matrix $C^k_j$ still has the same elements as $C_{ij}$, but represents an invertible map from $\mathbb{R}^{n+1}$ to itself. The linear map from $\mathbb{R}^{n+1}$ to $\mathbb{R}^{n+1*}$ that is represented by the matrix basically amounts to "transposition," since it maps a column vector $v^i$ to its transposed row vector $v_i = \delta_{ij} v^j$.

Since the effect of multiplying the resulting row vector $v_i$ times the original column vector $v^i$ is:

$$v_i v^i = \delta_{ij} v^i v^j, \tag{3.5}$$

which is the square of the Euclidian norm of **v**, one sees that the correlation on $\mathbb{R}P^n$ that is described by the matrix $\delta_{ij}$ is effectively defined by introducing a Euclidian scalar product on $\mathbb{R}^{n+1}$.

**4. The group of collineations of** $\mathbb{R}P^3$. Let $\mathcal{L}(3)$ be the set of all (projective) lines in $\mathbb{R}P^3$, as in the previous installments of this series of articles.

*a. Basic definition.* A *collineation* of a projective space $\mathbb{R}P^3$ is an invertible map from $\mathcal{L}(3)$ to itself; that is, it takes lines in $\mathbb{R}P^3$ to other such lines. Hence, from what we have just seen, there are basically two types of maps on $\mathbb{R}P^3$ that will take lines to lines: projective transformations and correlations.

Since collineations are invertible, the set $\mathcal{C}(3)$ of all collineations of $\mathbb{R}P^3$ forms a group under composition. One immediately sees that it must contain the group $PGL(3; \mathbb{R})$ of projective transformations. The remaining elements of $\mathcal{C}(3)$ then represent correlations, which define a set that is in one-to-one correspondence with $PGL(3; \mathbb{R})$ by choosing any correlation $[C]$ and multiplying each element $[L]$ of $PGL(3; \mathbb{R})$ by it. However, the coset $[C](PGL(3; \mathbb{R}))$ is not a subgroup, since the product of two correlations is not another correlation, and only one of the two disjoint subsets $PGL(3; \mathbb{R})$ and $[C](PGL(3; \mathbb{R}))$ will contain the identity.

*b. Representation of collineations on the Klein quadric.* Since the lines in $\mathbb{R}P^3$ can also be represented as points on the Klein quadric $\mathcal{K}$, one will naturally also ask about the nature projective transformations of $P\Lambda_2$ that will preserve that quadric; i.e., the invertible linear transformations of $\Lambda_2$ that take decomposable bivectors to decomposable bivectors. Those will then be the ones that preserve the scalar product:

$$<\mathbf{A}, \mathbf{B}> = V(\mathbf{A} \wedge \mathbf{B}) = \#\mathbf{A}(\mathbf{B}), \tag{4.1}$$

and one then sees that the group of invertible linear transformations of $\Lambda_2$ that preserve the scalar product that defines the Klein quadric will be isomorphic to $O(3, 3)$.

Like any orthogonal transformation – i.e., regardless of dimension or signature type – an element $A \in O(3, 3)$ will satisfy:

$$\det A = \pm 1. \tag{4.2}$$

An example of an element of $O(3, 3)$ that has a negative determinant is the involution $P: \Lambda_2 \to \Lambda_2$ whose matrix with respect to $\mathbf{E}_I$ is:

$$P_J^I = \begin{bmatrix} 0 & \delta_j^i \\ \hline \delta_j^i & 0 \end{bmatrix}. \quad (4.3)$$

The effect of this involution is to switch the subspace of elements the form $\mathbf{e}_0 \wedge \mathbf{a}$ with the subspace of elements of the form $\mathbf{a} \wedge \mathbf{b}$; i.e., to switch lines *to* infinity with lines *at* infinity. Thus, one can subdivide $O(3, 3)$ into the identity subgroup $SO(3, 3)$ and a group manifold that is diffeomorphic to it that can be obtained applying $P$ to the elements of $SO(3, 3)$. It is important to note that this involution $P$ is defined only when one chooses a basis for $\Lambda_2$, since the linear isomorphism of the vector spaces $\Lambda_2^{Re}$ and $\Lambda_2^{Im}$ is not canonical, any more than the decomposition of $\Lambda_2$ into $\Lambda_2^{Re} \oplus \Lambda_2^{Im}$ is canonical.

One can introduce a volume element on $\Lambda_2$ – i.e., a non-zero 6-form – by means of the reciprocal coframe $\{E^I, I = 1, \ldots, 6\}$ on $\Lambda^2$ (so $E^I(\mathbf{E}_J) = \delta_J^I$):

$$V^\wedge = E^1 \wedge \ldots \wedge E^6 = \frac{1}{6!} \varepsilon_{I_1 \cdots I_6} E^{I_1} \wedge \cdots \wedge E^{I_6}. \quad (4.4)$$

The effect of a linear transformation $A$ of $\Lambda_2$ on $V^\wedge$ is:

$$\frac{1}{6!} \varepsilon_{I_1 \cdots I_6} A_{J_1}^{I_1} E^{J_1} \wedge \cdots \wedge A_{J_6}^{I_6} E^{J_6} = \det [A] \, V^\wedge. \quad (4.5)$$

Since $\det [P] = -1$, one then sees that the involution $P$ has the effect of inverting the sign of this volume element; i.e., changing the orientation of the frame $\mathbf{E}_I$.

One can represent $PGL(3; \mathbb{R})$ in $PGL(P\Lambda_2)$ by means of the *exterior product representation*. That is, if $A \in GL(4; \mathbb{R})$ takes any vector $\mathbf{a} \in \mathbb{R}^4$ to $A(\mathbf{a})$ then $D(A)$ will take any bivector $\mathbf{a} \wedge \mathbf{b}$ to:

$$D(A)(\mathbf{a} \wedge \mathbf{b}) = A(\mathbf{a}) \wedge A(\mathbf{b}). \quad (4.6)$$

One can then extend this to the rest of $\Lambda_2$ by linearity, although for our purposes this will not be necessary, since we will mostly be interested in the decomposable case.

The representation $[D]$: $PGL(3; \mathbb{R}) \to PGL(P\Lambda_2)$ then takes $[\mathbf{a} \wedge \mathbf{b}] \mapsto [A(\mathbf{a}) \wedge A(\mathbf{b})]$. Since $A$ is invertible, $A(\mathbf{a}) \wedge A(\mathbf{b}) = 0$ iff $\mathbf{a} \wedge \mathbf{b} = 0$. Thus, the invertible linear map $D(A)$: $\Lambda_2 \to \Lambda_2$ that takes every $\mathbf{a} \wedge \mathbf{b}$ to $A(\mathbf{a}) \wedge A(\mathbf{b})$ (and gets extended by linearity) will take points on the quadric $\mathbf{B} \wedge \mathbf{B} = 0$ to other such points. One also finds that $\ker D = I$, since $D(A)(\mathbf{B}) = 0$ for all $\mathbf{B}$ iff $A(\mathbf{a}) \wedge A(\mathbf{b}) = 0$ for all $\mathbf{a}, \mathbf{b} \in \mathbb{R}^4$ iff $A(\mathbf{a})$ is collinear to $A(\mathbf{b})$ for all $\mathbf{a}, \mathbf{b} \in \mathbb{R}^4$ iff the image of $\mathbb{R}^4$ under $A$ is a line. This would contradict the invertibility of $A$, so the representation is faithful.

The fact that every element of *SO*(3, 3) can be expressed in the form of the tensor product representation of *PGL*(3; $\mathbb{R}$) follows from the fact that every point of $\mathcal{K}$ represents some line in $\mathbb{R}\mathrm{P}^3$. Hence, the image of *PGL*(3; $\mathbb{R}$) is *SO*(3, 3)

That then gives us the proof of a theorem that was discussed by Van der Waerden [**4**] that says how the collineations of $\mathbb{R}\mathrm{P}^3$ that do not permute lines at infinity with lines to infinity relate to the orthogonal transformations of $\Lambda_2$ :

**Theorem**. *One can define a group isomorphism* [*D*]: *PGL*(3; $\mathbb{R}$) $\cong$ *SO*(3, 3) *by means of the exterior product representation*.

As for the correlations, from the discussion above, we have:

**Theorem:** *The correlations of* $\mathbb{R}\mathrm{P}^3$ *are in one-to-one correspondence with the coset of SO*(3, 3) *under multiplication by* [*P*].

This gives the:

**Corollary:** $\mathcal{C}(3) \cong O(3, 3)$.

An interesting aspect of the isomorphism of the group of collineations of $\mathbb{R}\mathrm{P}^3$ with *O*(3, 3) is the fact that the former group does not depend upon the introduction of any scalar product on $\mathbb{R}\mathrm{P}^3$, and is therefore maximally general to three-dimensional projective geometry, while the latter group depends explicitly upon a scalar product for its definition, although it is a scalar product on a higher dimensional vector space than the one that covers $\mathbb{R}\mathrm{P}^3$. Hence, projective geometry can still be related to metric geometry even when one does not introduce a projective metric.

There is another situation in which a linear group of purely projective significance can be related to an orthogonal group that has great physical significance, namely, the case of *SL*(2; $\mathbb{C}$), which is the double covering group of the proper, orthochronous, Lorentz group $SO_0$(3, 1); i.e., its identity component.

Projectively speaking, *SL*(2; $\mathbb{C}$) is isomorphic to the group *PGL*(1; $\mathbb{C}$) of projective transformations of the complex projective line $\mathbb{C}\mathrm{P}^1$, which is diffeomorphic to the real 2-sphere as a manifold. Note that since the removal of the origin does not disconnect the complex plane $\mathbb{C}$, it is no longer necessary to mod out $\mathbb{Z}_2$ in order to get from *GL*(2; $\mathbb{C}$) to *PGL*(1; $\mathbb{C}$) $\cong$ *SL*(2; $\mathbb{C}$), as it is in the real case. One sees that all that is necessary in order to define $SL(2; \mathbb{C})$ is a volume element and a complex structure on $\mathbb{R}^4 = \mathbb{R}^2 \times \mathbb{R}^2$ that would make it $\mathbb{C}^2$.

Thus, the purely projective group $SL(2; \mathbb{C})$ is closely related to the orthogonal group of a space that is given the Minkowski scalar product.

**5. Reductions of the group of collineations of $\mathbb{R}P^3$.** Typically, the reduction of the linear group for some vector space or the group of collineations of some projective space comes about by introducing a function or tensor on that space and looking at the transformations that preserve that function or tensor. Since the two fundamental geometric objects of pre-metric electromagnetism are the electromagnetic constitutive tensor field and its associated dispersion law, which are then a fundamental tensor field on spacetime and a function on the cotangent bundle, we shall discuss the nature of the reductions of the group of collineations of $\mathbb{R}P^3$ that preserve those structures, first from a physical standpoint and then from a mathematical one.

The reader should be warned that due to the open-ended nature of the possible forms that electromagnetic constitutive laws and their associated dispersion laws might take, the discussion that follows must have and unavoidably general character to it.

*a. Collineations that preserve a constitutive law.* In order for an electromagnetic constitutive law to define an actual tensor field on spacetime $M$, one must already restrict oneself to electromagnetically linear, non-dispersive media, although it is not necessary for the medium to be electromagnetically homogeneous or isotropic. At each point $x \in M$, the electromagnetic constitutive law for the medium will then define a linear isomorphism $C(x) : \Lambda_x^2 \to \Lambda_{2, x}$, $F_x \mapsto \mathfrak{H}_x$, with

$$\mathfrak{H}_x = C(x)(F_x) . \tag{5.1}$$

In local components, it will then take the form:

$$\mathfrak{H}^{\mu\nu}(x) = C^{\mu\nu\kappa\lambda}(x) F_{\kappa\lambda}(x). \tag{5.2}$$

Hence, one can think of an electromagnetic constitutive law for an electromagnetically linear, non-dispersive medium as a fourth-rank, totally-contravariant tensor field:

$$C = C^{\mu\nu\kappa\lambda} (\partial_\mu \wedge \partial_\nu) \otimes (\partial_\kappa \wedge \partial_\lambda) . \tag{5.3}$$

It is clear from the fact that $C$ takes anti-symmetric, second-rank tensors to anti-symmetric, second-rank tensors that it must be anti-symmetric in its first and last pair of indices. However, it does not have to have any specific symmetry with respect to the permutation of those two index pairs, although one can decompose $C$ into a sum of three parts, the first of which $C_0$ is symmetric and does not include a part that is proportional to the volume element *V*, the second of which $C_+$ is anti-symmetric, and the third of which is proportional to *V*. (As mentioned previously, these components are called the *principal* part, the *skewon* part, and the *axion* part of *C*.)

In order to discuss reductions of the group of collineations of $\mathbb{R}P^3$, we now essentially restrict the scope of the discussion to the tangent and cotangent spaces to *M*, so we drop any explicit reference to the point *x*. We further replace the six-dimensional, real vector space $\Lambda_2$ with $\mathbb{R}^6$ and $\Lambda^2$ with $\mathbb{R}^{6*}$ by the introduction of a linear frame $\{\mathbf{E}_I, I = 1, \ldots, 6\}$ on $\Lambda_2$, which might be defined by a linear frame $\{\mathbf{e}_\mu, \mu = 0, \ldots, 3\}$ on $\mathbb{R}^6$.

The tensor *C* then takes the form:

$$C = C_{IJ} E^I \otimes E^J, \tag{5.4}$$

where $\{E^I, I = 1, \ldots, 6\}$ is the reciprocal coframe field to $\mathbf{E}_I$.

As mentioned before, such a tensor is then associated with a linear isomorphism $C : \mathbb{R}^{6*} \to \mathbb{R}^6$ and a correlation $[C] : \mathbb{R}P^5 \to \mathbb{R}P^{5*}$, which takes points in $\mathbb{R}P^5$ to hyperplanes in that space. It can also be defined as a bilinear functional $C(A, B)$ on $\mathbb{R}^{6*}$. Like any other correlation, *C* does not have to have any symmetry as a bilinear functional, although the principal part of *C* and its axion parts will both be symmetric, while the axion part will be anti-symmetric.

Naively, one might start with the notion of invertible linear transformations of $\mathbb{R}^{6*}$ that preserve *C*, i.e., $T \in GL(6, \mathbb{R})$ such that:

$$C(TA, TB) = C(A, B), \qquad \text{for all } A, B \text{ in } \mathbb{R}^{6*}. \tag{5.5}$$

One could also express this in its component form as:

$$T^I_K T^J_L C^{KL} = C^{IJ} \qquad \text{or} \qquad T^T C T = C. \tag{5.6}$$

However, one might rapidly find that there might not be any such transformations, or at least very many of them besides the identity transformation *I*.

Basically, one is dealing with the tautology: The less symmetric a geometric object is, the less symmetry transformations that it will admit. Think of an elastic two-dimensional sphere in a three-dimensional vector space. If one pushes down on its North pole then it will become an oblate spheroid, which will still be a surface of revolution, so the symmetry group has been reduced from $O(3)$ to $O(2)$. If one further stretches the surface in both directions along a diameter of the equator then the resulting surface will be a general ellipsoid, which has only finite reflection symmetries about its three axes, so the symmetry group would be $(\mathbb{Z}_2)^3$.

However, even in the last case, if one chooses a frame that is orthonormal for the scalar product that the ellipsoid defines then the ellipsoid will become a sphere in that frame. Thus, one typically thinks of the symmetry of a quadric (hyper)surface as something that relates to the signature type of its coefficient matrices as a quadratic form, not to the particular matrix in a particular frame.

In the present case of the correlation $C$, only the symmetric part will define a quadratic form on $\mathbb{R}^{6*}$, and since the volume element $V$ already defines its own quadric (viz., the Klein quadric), one should probably just consider the quadric that is defined by the principal part $C_0$ by itself.

As for the skewon part of $C$, it would define a symplectic structure on $\mathbb{R}^{6*}$, if it were non-degenerate. (Of course, when $C$ is composed of a sum of three bilinear forms, the individual forms do not all have to be non-degenerate, but only their sum.) Since, by Darboux's theorem, all symplectic structures on the same even-dimensional real vector space will be equivalent, there will be essentially just one type of non-degenerate skewon.

One can get some idea of the typical signature type of the principal part of $C$ by looking at the most elementary electromagnetic constitutive laws, such as the classical vacuum and the constitutive law that goes with the Hodge * for a Lorentzian metric, namely, the map from $\Lambda^2$ to $\Lambda_2$ that raises both indices of the components. One finds that both of the latter constitutive laws define scalar products that have a signature type of $(-1, -1, -1, +1, +1, +1)$, as does the scalar product that $V$ defines. However, one should be cautioned that the two scalar products that are defined by $C_0$ and $V$ will not generally be diagonalized in the same linear frames.

Hence, one suspects that even for the more general constitutive laws, the collineations that preserve the quadric of $C_0$ will relate to $O(3, 3)$, or its conformal group, since the quadric is defined by a homogeneous equation, namely:

$$C_0(F, F) = 0. \qquad (5.7)$$

This equation becomes the generalization of the elementary condition $\mathbf{E}^2 - \mathbf{H}^2 = 0$ that defines one of the two necessary (but not sufficient) conditions for an electromagnetic field to wave-like in the classical vacuum. The other one – viz., $\mathbf{E} \cdot \mathbf{H} = 0$ – is equivalent to the condition $V(F, F) = 0$ that puts the 2-form $F$ on the (dual) Klein quadric, and implies that $F$ must be decomposable, such as $k \wedge a$, and this represent a line in $\mathbb{R}P^3$.

Of course, when one passes from $\mathbb{R}^{6*}$ to $\mathbb{R}P^{5*}$ the only change in the group $O(3, 3)$ will be to restrict to $SO(3, 3)$, since $PGL(5, \mathbb{R})$ is isomorphic to $SL(6, \mathbb{R}) / \mathbb{Z}_2$.

*b. Collineations that preserve a dispersion law.* As discussed above, a dispersion law often takes the form of a homogeneous polynomial equation on covectors, which take the form of the frequency-wave number 1-form $k = \omega\, dt - k_i\, dx^i$:

$$D[k] = 0. \qquad (5.8)$$

One should not assume that the inhomogeneous case is physically uninteresting, since it includes not only the dispersion law of massive free particles, but also some types of plasmas.

The degree of the polynomial $D$ will be either two or four, where four is the general case and two is a degenerate, but popular, case. In particular, the Lorentzian metric $g$ is associated with the possibility that the electromagnetic constitutive law implies a dispersion polynomial of the form:

$$D[k] = (g(k, k))^2. \qquad (5.9)$$

Since the square does not change anything in the homogeneous equation, one can then reduce to the quadratic dispersion law:

$$g(k, k) = 0, \qquad (5.10)$$

which defines the light cones in the cotangent spaces of a Lorentzian manifold, and for an orthonormal frame they will look like:

$$\eta(k, k) = 0, \qquad \eta^{\mu\nu} = \text{diag}(-1, -1, -1, +1), \qquad (5.11)$$

where $\eta^{\mu\nu}$ is the component matrix of the Minkowski scalar product in an orthonormal coframe.

This last dispersion law is the one that takes the form:

$$\omega = \pm c\kappa, \qquad \kappa = \| k_i \, dx^i \|, \qquad (5.12)$$

where the spatial norm is the Euclidian one. This dispersion law is often referred to "linear," although in the present context that might be confusing, since we also think of it as quadratic. Another source of confusion is the fact that this dispersion law is associated with "absence of dispersion" in the sense that $c$ is only a constant function of $k$.

The next most general case beyond Lorentzian is the *bi-metric* case:

$$D[k] = g_1(k, k) \, g_2(k, k), \qquad (5.13)$$

where $g_1$ and $g_2$ typically both have a Lorentzian signature type. However, they are not simultaneously diagonalizable in the same orthonormal frame, since that would bring one back to the degenerate case of (5.9). This case is particularly relevant to the Heisenberg-Euler one-loop effective model for charged particles that move in external electromagnetic fields, for which $g_1$ and $g_2$ take the form of the Minkowski scalar product $\eta^{\mu\nu}$, perturbed by terms that are proportional to the energy-moment-stress tensor of the charge-external field system ([1]).

The most general case is a homogeneous quartic that might still exhibit *birefringence* (or double refraction). The way that one gets that phenomenon as a consequence of (5.8) is by fixing the direction of propagation of an electromagnetic wave (i.e., $k_i \, dx^i$), so the dispersion law becomes a homogeneous quartic polynomial in $\omega$, which usually takes the form of a homogeneous quadratic equation in $\omega^2$. Thus, there will generally be two distinct roots for $\omega^2$ and since the square roots of $\omega^2$ are $\pm \omega$, and the sign only describes the direction of propagation, one can get two generally distinct velocities of propagation in the given spatial direction from either the phase velocity or group velocity that is as associated with $k$ and $D$. Although it is not at all obvious from the present discussion, the way that nature chooses one or the other velocity is by way of the state of polarization of the electromagnetic wave, which we have not introduced explicitly, as of yet.

---

[1] For a more detailed discussion of the geometry of electromagnetic constitutive laws and dispersion laws, one might confer the author's book [**5**].

The most natural (if naïve) definition of the (linear) symmetry group of a dispersion law is the set of all invertible linear maps of $\mathbb{R}^{4*}$ that preserve the *equation* (5.8); i.e., all $T \in GL(4; \mathbb{R})$ such that:

$$D[T(k)] = 0 \quad \text{iff} \quad D[k] = 0. \tag{5.14}$$

Of course, if one wishes to start with the established facts of classical electromagnetism in its metric formulation then one must also expand the scope of the transformations from merely invertible linear ones to diffeomorphisms. One would then get the Lorentzian conformal group for Minkowski space as the symmetry group for the Lorentzian dispersion law (5.11). The linear subgroup of that group consists of the Lorentz group, together with the homotheties (i.e., non-zero scalar multiplications). The nonlinear transformations that it leaves out are the four-dimensional translations and the inversions through light cones, which relate to the transformation to constantly-accelerated moving frames.

Now, in the bi-metric case, since one can never simultaneously diagonalize the two metrics (or else it would be a Lorentzian dispersion law), one will not generally have very many non-trivial linear transformations that simultaneously preserve both light cones. For instance, if one of them is spatially spherical, while the other one is spatially an oblate spheroid, their common symmetries would belong to $SO(2)$. However, if one is more concerned with *motions*, in general, that *symmetries*, which are a special type of motion, then one might think of the transformations that move covectors around the two light cones as involving two distinct representations of the conformal Lorentz group, along with a choice of diffeomorphism that that takes one light cone to the other one. It might also make a difference whether the two light cones do or do not intersect.

In the general case, one expects that the wave quartic might behave more like the Fresnel wave surface for biaxial optical media or the Kummer surface, which is a generalization of the latter surface ([1]). Since the linear symmetries of such a quartic hypersurface define only a finite group of transformations, once again, one might consider the diffeomorphisms that take points of the hypersurface to other such points; i.e., motions of wave vectors.

*c. Collineations that preserve an almost-complex structure.* Since we already introduced the concept of an almost-complex structure in Part II of this series [**2**], we shall only briefly summarize the facts that relate to the present discussion, while expanding upon some points that were passed over previously.

---

([1]) For a thorough discussion of the physical issues that are involved with the generalization from the Fresnel wave surface to the Kummer surface (and possibly beyond), one should confer [**6**].

An *almost-complex structure* on an even-dimensional, real vector space $V$ is a (real) linear isomorphism $* : V \to V$ such that:

$$*^2 = -I. \tag{5.15}$$

(The fact that $V$ must be even-dimensional follows from the fact that the real dimension of $\mathbb{C}^n$ is $2n$.)

For instance, if $V = \mathbb{R}^2$ then the isomorphism $i : \mathbb{R}^2 \to \mathbb{R}^2$, $(x, y) \mapsto (-y, x)$ will have this property. The matrix of this map relative to the canonical frame on $\mathbb{R}^2$ is then:

$$[i] = \begin{bmatrix} 0 & -1 \\ 1 & 0 \end{bmatrix}, \tag{5.16}$$

which also represents a counter-clockwise rotation through a right angle in the plane.

It is important to notice here that the reason that complex or almost-complex structures are often associated with orthogonal structures in some form is generally based upon this key fact about the imaginary $i$ that it basically represents a rotation in the plane. Thus, it is not as surprising now that the group $SL(2; \mathbb{C})$, which involves only a complex structure (and a volume element) for its definition, should be related to $SO(3, 1)$, which involves an orthogonal structure.

When one is given an almost-complex structure on an even-dimensional, real vector space $V$, one can then define a complex structure, as well, by simply defining complex scalar multiplication on $V$. In order to do that, one starts with the definition of multiplication by $i$:

$$i\mathbf{v} \equiv *\mathbf{v} \tag{5.17}$$

and extends to the other complex scalars by way of complex linearity:

$$(u + iv)\,\mathbf{v} = u\,\mathbf{v} + v*\mathbf{v}. \tag{5.18}$$

Since $*^2 = -I$, the eigenvalues of $*$ will be $\pm i$. Of course, in order to define eigenvectors that would go with them, one must regard $V$ as a complex vector space, not a real one. The eigenvector equation would then take the form $*\mathbf{v} = \pm i\,\mathbf{v}$, but if one recalls the definition (5.17), one would see that choosing the positive sign will make *every* vector in $V$ an eigenvector and choosing the negative sign would simply define a different complex structure on $V$, which is often referred to as the *opposite* complex structure.

One can still polarize $V$ using $*$:

$$\mathbf{v} = \mathbf{v}_+ + \mathbf{v}_-, \tag{5.19}$$

with

$$\mathbf{v}_\pm = \tfrac{1}{2}(\mathbf{v} \pm {}^*\mathbf{v}). \qquad (5.20)$$

Thus, one will have:

$$^*\mathbf{v}_\pm = \tfrac{1}{2}({}^*\mathbf{v} \mp \mathbf{v}) = \mp\, \mathbf{v}_\mp ; \qquad (5.21)$$

i.e.:

$$^*\mathbf{v}_+ = -\mathbf{v}_-, \qquad ^*\mathbf{v}_- = \mathbf{v}_+. \qquad (5.22)$$

This gives us a direct-sum decomposition $V = V_+ \oplus V_-$ into (real) isomorphic subspaces that behave somewhat like real and imaginary subspaces, although with the opposite sign. Thus, if the dimension of $V$ is $2n$, and we have a real-linear basis $\{\mathbf{e}_i, i = 1, \ldots, n\}$ for the "real" subspace $V_-$ then we can also define a real-linear basis for the "imaginary" subspace $V_+$ by way of $^*\mathbf{e}_i$, and $\mathbf{e}_i$ will also define a complex-linear basis for $V$. Thus, any $\mathbf{v}$ can be expressed in the form:

$$\mathbf{v} = (v^i + i\, w^i)\, \mathbf{e}_i = v^i\, \mathbf{e}_i + w^i\, {}^*\mathbf{e}_i ; \qquad (5.23)$$

so

$$^*\mathbf{v} = -w^i\, \mathbf{e}_i + v^i\, {}^*\mathbf{e}_i , \qquad (5.24)$$

which makes:

$$\mathbf{v}_\pm = \tfrac{1}{2}[(v^i \mp w^i)\, \mathbf{e}_i + (w^i \pm v^i)\, {}^*\mathbf{e}_i]. \qquad (5.25)$$

The even-dimensional vector spaces of interest to electromagnetism would be $\Lambda_2 = \Lambda_2 \mathbb{R}^4$ and its dual vector space $\Lambda^2 = \Lambda^2 \mathbb{R}^4 \cong (\Lambda_2 \mathbb{R}^4)^*$, which are six-real-dimensional. If $\{\mathbf{e}_\mu, \mu = 0, \ldots, 3\}$ is a basis for $\mathbb{R}^4$ then one can define a basis for $\Lambda_2$ by way of:

$$\mathbf{E}_i = \mathbf{e}_0 \wedge \mathbf{e}_i , \qquad \mathbf{E}_{i+3} = \varepsilon_{ijk}\, \mathbf{e}_j \wedge \mathbf{e}_k . \qquad (5.26)$$

This defines a direct-sum decomposition $\Lambda_2 = \Lambda_2^{\mathrm{Re}} \oplus \Lambda_2^{\mathrm{Im}}$ according to the subspaces that are spanned by $\mathbf{E}_i$ and $\mathbf{E}_{i+3}$, respectively.

Note that $\Lambda_2^{\mathrm{Re}}$ is the image of $\mathbb{R}^3$ under the linear injection $\mathbb{R}^3 \to \Lambda_2$, $\mathbf{v} \mapsto \mathbf{e}_0 \wedge \mathbf{v}$ and $\Lambda_2^{\mathrm{Im}}$ is isomorphic to $\Lambda_2 \mathbb{R}^3$.

Using this basis, we can then define an almost-complex structure on $\Lambda_2$ by way of:

$$^*\mathbf{E}_i = \mathbf{E}_{i+3}, \qquad ^*\mathbf{E}_{i+3} = -\mathbf{E}_i . \qquad (5.27)$$

This means that one can also write the real-linear basis for $\Lambda_2$ in the form of $\{\mathbf{E}_i, {}^*\mathbf{E}_i\}$, so $\{\mathbf{E}_i\}$ will define a complex-linear basis for it as a three-dimensional complex vector space. Thus, one can use this basis to show that as a complex vector space $\Lambda_2 \cong_{\mathbb{C}} \mathbb{C}^3$, where the isomorphism will then take the form $\mathbb{C}^3 \to \Lambda_2$, $(A^i + iB^i) \mapsto A^i\, \mathbf{E}_i + B^i\, {}^*\mathbf{E}_i$.

A useful property of the isomorphism $*$ is:

**Theorem:**
$$\langle \mathbf{E}_I, *\mathbf{E}_J \rangle = \langle *\mathbf{E}_I, \mathbf{E}_J \rangle.$$

**Proof:**

From (5.27), one computes:

$$\begin{aligned}
\langle \mathbf{E}_i, *\mathbf{E}_j \rangle &= \langle \mathbf{E}_i, \mathbf{E}_{j+3} \rangle = -\langle *\mathbf{E}_{i+3}, \mathbf{E}_j \rangle = \delta_{ij}, \\
\langle \mathbf{E}_i, *\mathbf{E}_{j+3} \rangle &= -\langle \mathbf{E}_i, \mathbf{E}_j \rangle = \langle *\mathbf{E}_{i+3}, \mathbf{E}_j \rangle = 0, \\
\langle \mathbf{E}_{i+3}, *\mathbf{E}_j \rangle &= \langle \mathbf{E}_i, \mathbf{E}_{j+3} \rangle = -\langle *\mathbf{E}_{i+3}, \mathbf{E}_{j+3} \rangle = 0, \\
\langle \mathbf{E}_{i+3}, *\mathbf{E}_{j+3} \rangle &= -\langle \mathbf{E}_{i+3}, \mathbf{E}_j \rangle = -\langle *\mathbf{E}_i, \mathbf{E}_j \rangle = -\delta_{ij},
\end{aligned}$$

and matches up corresponding expressions.

This has the immediate, obvious:

**Corollaries:**

1. $\langle \mathbf{A}, *\mathbf{B} \rangle = \langle *\mathbf{A}, \mathbf{B} \rangle$.
2. $\langle *\mathbf{A}, *\mathbf{B} \rangle = -\langle \mathbf{A}, \mathbf{B} \rangle$.
3. * maps decomposable bivectors to decomposable bivectors.
4. * preserves the Klein quadric.

The first of these says that the map * is self-adjoint with respect to the scalar product $\langle .,. \rangle$, while the second one says that it is "anti-orthogonal;" i.e., it preserves the scalar product, but inverts the sign.

The presence of an almost-complex structure * on $\Lambda_2$, and thus a complex structure, will allow us to define a subgroup of $GL(\Lambda_2)$ that is isomorphic to $GL(3; \mathbb{C})$. One basically restricts oneself to all invertible linear transformations $L: \Lambda_2 \to \Lambda_2$ that commute with *:

$$L* = *L. \qquad (5.28)$$

If one defines a real frame of the form $\{\mathbf{E}_i, *\mathbf{E}_i\}$ on $\Lambda_2$ then if the matrices of $L$ and * take the form:

$$[L] = \begin{bmatrix} A & B \\ \hline C & D \end{bmatrix}, \qquad [*] = \begin{bmatrix} 0 & -I \\ \hline I & 0 \end{bmatrix} \qquad (5.29)$$

then when one does the multiplications in (5.28), one will find that this will demand that $A = D$ and $B = -C$, which means that the matrix of $L$ must take the form:

$$[L] = \begin{bmatrix} A & -B \\ \hline B & A \end{bmatrix} = \begin{bmatrix} A & 0 \\ \hline 0 & A \end{bmatrix} + \begin{bmatrix} 0 & -B \\ \hline B & 0 \end{bmatrix}, \qquad (5.30)$$

which amounts to the "real + imaginary" form of a complex matrix. The isomorphism with $GL(3; \mathbb{C})$ will then take the form of the associations:

$$\begin{bmatrix} A & 0 \\ 0 & A \end{bmatrix} \mapsto [A], \qquad \begin{bmatrix} 0 & -B \\ B & 0 \end{bmatrix} \mapsto i\,[B].$$

Note that there is an important difference between the reduction of $GL(\Lambda_2)$ to $SL(\Lambda_2)$ and the reduction of $GL(3; \mathbb{C})$ to $SL(3; \mathbb{C})$:

In the former (real) case, since $\Lambda_2$ has a real dimension of six as a vector space, in order to reduce from $GL(\Lambda_2)$ to $SL(\Lambda_2)$ one must define a real 6-form to be the volume element on the vector space and a corresponding real determinant on 6×6 real matrices. The resulting subgroup will have a real dimension of 35.

In the latter (complex) case, $\mathbb{C}^3$ has a complex dimension of three, so the volume element will be a complex 3-form, and the corresponding determinant will be a complex function of 3×3 complex matrices. Hence, the resulting subgroup will have a complex dimension of eight, which gives a real dimension of *sixteen*. Thus, the imposition of a complex structure on $\Lambda_2$ has reduced the possibilities considerably as far as linear transformations are concerned.

The natural volume element to define on $\mathbb{C}^3$ is defined by the canonical basis $\{\delta_i, i = 1, 2, 3\}$ and its reciprocal basis $\{\delta^i, i = 1, 2, 3\}$:

$$V_\mathbb{C} = \delta^1 \wedge \delta^2 \wedge \delta^3 = \frac{1}{3!}\varepsilon_{ijk}\delta^i \wedge \delta^j \wedge \delta^k. \tag{5.31}$$

Although $\delta_i$ corresponds to $\mathbf{E}_i$ under the $\mathbb{C}$-isomorphism of $\Lambda_2$ with $\mathbb{C}^3$, and $*\mathbf{E}_i$ corresponds to $i\delta_i$, one cannot relate $V_\mathbb{C}$ to $V$ by simply performing the same exterior multiplication of all six real basis elements, since $*\mathbf{E}_1 \wedge *\mathbf{E}_2 \wedge *\mathbf{E}_3$ will correspond to $-iV_\mathbb{C}$, and since that is not $\mathbb{C}$-linearly independent of $V_\mathbb{C}$, the product $V_\mathbb{C} \wedge (-iV_\mathbb{C})$ will have to vanish.

*d. Complex orthogonal transformations.* As observed previously [2], when $\Lambda_2$ has been given an almost-complex structure *, and thus a complex structure, the real scalar product $\langle \mathbf{A}, \mathbf{B}\rangle = V(\mathbf{A} \wedge \mathbf{B})$ that is defined by $V$ can be extended to a complex scalar product. One first defines:

$$(\mathbf{A}, \mathbf{B}) \equiv \langle \mathbf{A}, *\mathbf{B}\rangle \tag{5.32}$$

and then defines:

$$\langle \mathbf{A}, \mathbf{B}\rangle_\mathbb{C} = (\mathbf{A}, \mathbf{B}) + i\langle \mathbf{A}, \mathbf{B}\rangle. \tag{5.33}$$

In the definition (5.32), we have implicitly used the *self-adjointness* of * with respect to the scalar product $\langle .,. \rangle$:

$$\langle \mathbf{A}, *\mathbf{B} \rangle = \langle *\mathbf{A}, \mathbf{B} \rangle \tag{5.34}$$

that we mentioned above in a corollary.

The presence of this complex-Euclidian scalar product (5.33) on $\Lambda_2$ also allows one to reduce the group $SO(3, 3)$ of invertible real-linear transformations of $\Lambda_2$ that preserve scalar product $\langle ., . \rangle$ that defines the Klein quadric, along with a volume element on $\Lambda_2$, to the group $SO(3; \mathbb{C})$ of invertible complex-linear transformations of $\Lambda_2$ that also preserve the scalar product $(., .)$, as well. Similarly, one could reduce the $GL(3; \mathbb{C})$ subgroup of $GL(\Lambda_2)$ to $SO(3; \mathbb{C})$ by first introducing the complex orthogonal structure and then a volume element.

The group $SO(3; \mathbb{C})$ is intimately-related to special relativity, which is based in the propagation of electromagnetic waves, by the fact that it is isomorphic to $SL(2; \mathbb{C})$, which is the double covering of the proper, orthochronous Lorentz group $SO_0(3, 1)$. Hence, one can think of three distinct geometric contexts in which to discuss special relativity, according to which representation of the basic group one chooses:

1. Real, four-dimensional Minkowski space ($\mathbb{R}^4$, $\eta_{\mu\nu}$), which corresponds to $SO_0(3, 1)$. Hence, one deals with real, hyperbolic geometry.

2. $\mathbb{C}P^1$, which corresponds to the group $SL(2; \mathbb{C})$. Here, one deals with complex projective geometry for the complex projective line, which is the same as the real 2-sphere as a real manifold.

3. Complex, three-dimensional Euclidian space, which corresponds to the group $SO(3; \mathbb{C})$. In this context, one is then concerned with a special class of lines in $\mathbb{R}P^3$, namely, the ones whose bivectors satisfy:

$$(\mathbf{A}, \mathbf{A}) = 0, \tag{5.35}$$

in addition to the constraint that makes them represent lines:

$$\langle \mathbf{A}, \mathbf{A} \rangle = 0. \tag{5.36}$$

If $\Lambda^2$ is given an almost-complex structure * then, since the complex-Euclidian structure on $\Lambda^2$ is defined by the scalar products that are associated with $\mathcal{K}$ and $C_s$, one will see that the linear transformations of $\Lambda^2$ that take the electromagnetic wave planes (or their corresponding lines in $\mathbb{R}P^3$) to other such planes will belong to an $SO(3; \mathbb{C})$ subgroup of $SO(3, 3)$. Hence, we see that the reduction in symmetry from collineations to Lorentz transformations is associated simply with a particular type of electromagnetic constitutive law.

It is important to note that not all linear, non-dispersive, electromagnetic, constitutive laws will define almost-complex structures. For more details on this topic, one should confer the author's paper [**7**].

**6. Contact transformations.** The subject of contact geometry, which can be regarded as an extension of projective geometry, and contact transformations is really quite vast in scope (see, e.g., the tomely treatise of Lie and Scheffers [**8**]). Consequently, all that we shall include here are a few perfunctory remarks concerning how contact transformations relate to the subjects of line geometry and electromagnetism.

A *contact element* is a (proper) linear subspace of a tangent space to a point $x$ in a differentiable manifold that is tangent to some submanifold of the same dimension that is defined in a neighborhood of $x$. In the case of four-dimensional manifolds, that would include lines, planes, and hyperplanes, which would then become points, lines, and planes in the projective spaces that the tangent spaces define (i.e., the *projectivized tangent bundle*). Since these tangent projective spaces are projectively equivalent to $\mathbb{R}P^3$, one sees that a line in that tangent projective space is already associated with both a point on the Klein quadric $\mathcal{K}$ in the space $P\Lambda_{2,x}$, and one on its dual $\mathcal{K}^*$ in $P\Lambda_x^2$, even in the absence of an electromagnetic field. Thus, a contact plane is also associated with corresponding points of $\mathcal{K}$ and $\mathcal{K}^*$.

As long as one has some way of giving physical meaning to planes in tangent spaces or lines in the tangent projective spaces, one can then define an immediate link between contact planes and line geometry, as we have been discussing it up to this point. In fact, that link is obvious in the context of the propagation of waves (of any sort) in three-dimensional spaces (i.e., four-dimensional spacetimes), since a differentiable wave surface will have a unique contact element at each point that will generally evolve in time in a manner that is constrained by a dispersion law. Indeed, the very shape of the instantaneous wave fronts (which is what one calls the intersection of the dispersion hypersurface with the "simultaneity hypersurfaces") is dictated by the nature of that dispersion law. More generally, one might consider self-intersecting surfaces, such as the Fresnel or Kummer wave surfaces, for which one would have a finite number of distinct contact planes at some (singular) points.

A *contact transformation* is a special type of map of the tangent bundle $T(M)$ to itself ([1]) that takes contact elements to contact elements. For instance, a diffeomorphism of $M$ will have invertible linear maps for its differential maps at each point, which will take tangent planes to tangent planes. However, it must also preserve the tangency of that plane with some associated locally-defined surface in $M$, which is a condition that is best treated in the language of jet manifolds.

One can then see that the time evolution of wave surfaces in space is associated with a corresponding one-parameter family of contact transformations. Hence, the group of motions that seems to be most intrinsic to wave motion in three-dimensional space is the group of contact transformations that pertains to planar contact elements. Unlike the conformal Lorentzian group, which is 15-dimensional, groups of contact transformations are typically infinite-dimensional. Indeed, since one defines such transformations most generally by integrating a system of partial differential equations, which usually produces local solutions (if at all), it is often best to think of contact transformations as a topic that is best treated in the language of "Lie pseudogroups." Of course, all of these topics are rapidly drifting far beyond the scope of the immediate discussion, which is line geometry and electromagnetism. (Although quite of a bit of the material in Lie and Scheffers [**8**] relates to the role of line geometry in the theory of contact transformations.)

---

([1]) It is much more geometrically intrinsic to define the space of contact planes as the manifold of "1-jets of local differentiable maps from $\mathbb{R}^2$ into $M$." For a clarification of this, one might confer the author's paper on pre-metric wave mechanics [**9**].

Since line geometry apparently has much to say about wave motion in its general context, as does contact geometry, it should then be clear that the three subjects would all relate to the electromagnetic waves in space, in particular.

**7. Summary.** When one looks at the (inverted) tree of subgroup inclusions that has the fifteen-dimensional group $PGL(3, \mathbb{R})$ of collineations of $\mathbb{R}P^3$, which take lines to lines, one sees that the appearance of the Lorentz group first comes about after one has introduced two crucial physical objects. Namely, one must introduce an electromagnetic constitutive law and its associated dispersion law, and then restrict oneself to a very specific class of constitutive laws that imply degenerate quartic dispersion laws that take the form of the squares of quadratic constitutive laws of Lorentzian type. Such constitutive laws will also define almost-complex structures on the bundle $\Lambda^2(M)$ of 2-forms on a four-dimensional spacetime manifold $M$, which allow one to reduce the group of collineations, first to a $GL(3; \mathbb{C})$ subgroup and then to $O(3, \mathbb{C})$. The introduction of an orientation on $\Lambda^2(M)$ then allows one to reduce to $SO(3; \mathbb{C})$, which is isomorphic to $SL(2; \mathbb{C})$, and thus defines another representation of the Lorentz group that pertains to complex relativity and the Riemann-Silberstein-Majorana-Oppenheimer representation of electromagnetic fields by sections of a complex vector bundle whose fibers have a complex dimension of three. (These sections are sometimes referred to as "3-spinors.")

One then sees a group-theoretic way by which the theory of relativity and its explanation for the presence of gravitation in the universe as being a consequence of a Lorentzian metric with a curved Levi-Civita connection becomes a corollary to a much broader picture of the geometry of spacetime as seen through the eyes of its electromagnetic structure. In particular, the Lorentzian metric is now a degenerate case of a more general dispersion law that might become non-degenerate if vacuum birefringence is associated with vacuum polarization, such as in the cloud that surrounds elementary charges, according to charge renormalization.

This suggests that one interesting class of problems in pre-metric electromagnetism might be that of finding electromagnetic constitutive laws that imply the most elementary solutions of the Einstein field equations beyond the Minkowski metric; one might expect such constitutive laws to be linear and isotropic, but inhomogeneous. (The Schwarzschild and Robertson-Walker metrics night make a good place to start.) This direction of research is, of course, closely related to the "complex relativity" approach to gravitation, which also happens to overlap with the representation of electromagnetism by 3-spinors. The author has taken some first steps in that direction (see, [**5**] and [**10**].) If one thinks of the principle of relativity (in any of its forms) as a principle of invariance under the action of a group of transformations then one sees that physically everything depends upon the geometric nature of the dispersion law, if not the constitutive law.

Of course, the most physically fundamental problem in all of this is that of reducing the generality of the physically-interesting constitutive laws and dispersion laws, by producing some classification of the physical possibilities that might imply corresponding line-geometric or group-theoretic equivalence classes. In particular, the transitions from the general homogeneous quartic polynomial to a product of Lorentzian factors and then to a square of one Lorentzian factor seem quite fundamental to the physics of electromagnetism.

Another topic of a group-theoretic nature that pertains to electromagnetism is that of the symmetries of the *equations* of pre-metric electromagnetism, which is in line with what Lie was doing when he first came up with his theory of continuous transformation groups, as well as what Bateman and Cunningham were doing in the name of electromagnetism. Once again, the fundamental issue is what sort of symmetries the constitutive law exhibits. The author has also taken some first steps in formulating that problem, as well (cf., [**11**]).

In summary, the geometry of electromagnetism seems to be that of Felix Klein, line geometry, and scalar products on the bundles of bivector fields and 2-forms, while the geometry of gravitation − namely, the geometry of Riemann and Lorentzian metrics on the tangent bundle − takes its place in the hierarchy of geometries as a degenerate case of a more general situation.

Line geometry and electromagnetism III: groups of transformations. 238.  S. Lie and G. Scheffers, *Berührungstransformationen*, Teubner, Leipzig, 1896; republished by A. M. S. Chelsea Publishing, Providence, R. I., 2005.
9.  D. H. Delphenich, "On the pre-metric foundations of wave mechanics I: massless waves," Ann. Phys. (Leipzig) **18** (2009), 206-230, and arXiv:0904.1960.
10. D. H. Delphenich, "A more direct representation for complex relativity," Ann. Phys. (Leipzig) **16** (1997), 615-639, and arXiv:0707.4681.
11. D. H. Delphenich, "Symmetries and pre-metric electromagnetism," Ann. Phys. (Leipzig) **14** (2005), 663-704, and gr-qc/0508035; this paper was based upon a talk that the author gave in Kiev in that year: "Symmetries of the equations of pre-metric electromagnetism," gr-qc/0512126.